\documentclass[aps,twocolumn,raggedbottom,prb,showpacs,nobalancelastpage,amssymb,groupedaddress]{revtex4}
\usepackage{graphicx}
\usepackage{amsmath}
\usepackage{amssymb}

\newcommand{\bra}[1]{\langle #1|}
\newcommand{\ket}[1]{|#1\rangle}

\def\Q{{\@QC Q}}
\def\C{{\@QC C}}
\def\@QC#1{\mathpalette{\setbox0=\hbox\bgroup$\rm}%
  {\egroup C$\egroup\rm\rlap{\kern0.4\wd0\vrule
  width 0.05\wd0 height 0.97\ht0 depth -0.01\ht0}%
  #1\bgroup}}

\newcommand{\eq}[2]{
\begin{equation}
\label{#1}
#2
\end{equation}}

\newcommand{\vm}{\vec{m}}
\newcommand{\vN}{\vec{N}}

\newcommand{\da}{a^{\dagger}}
\newcommand{\Psid}{\Psi^{\dagger}}

\begin{document}
\title{Exchange effects in spin polarized transport through carbon nanotube quantum dots}
%\title{Detector induced resonances in a driven solid-state qubit}
\author{Christoph Schenke, Sonja Koller, Leonhard Mayrhofer, and Milena Grifoni}
\affiliation{Institut f\"{u}r Theoretische Physik, Universit\"at
Regensburg, 93035 Regensburg, Germany}
\date{\today}

\begin{abstract}
We investigate linear and nonlinear transport across single-walled carbon nanotube quantum dots weakly coupled to spin-polarized leads. We consider metallic tubes of finite length and small diameter, where not only forward scattering contributions of the Coulomb potential, but also short-ranged processes play an important role. In particular, they induce exchange effects leading for electron fillings $4n+2$ either to a non-degenerate groundstate of spin $S=0$ or to a triplet groundstate. In the linear regime we present analytical results for the conductance - for both the $S=0$ and the triplet groundstate - and demonstrate that an external magnetic field is crucial to reveal the spin nature of the groundstates. In the nonlinear regime we show stability diagrams that clearly distinguish between the different groundstates. We observe a negative differential conductance (NDC) effect in the $S=0$ groundstate for antiparallel lead magnetization. In presence of an external magnetic field, spin blockade effects can be detected, again leading to NDC effects for both groundstates.
\end{abstract}
 \maketitle

\section{Introduction}

Since their discovery by S. Iijima and T. Ichihashi \cite{Iijima} in 1993, single-walled carbon nanotubes (SWNTs) have attracted attention due to their remarkable electronic and mechanical properties \cite{Saito,Loiseau}. At low energies, they represent an almost perfect realization of a one-dimensional (1D) system of interacting electrons with an additional orbital degree of freedom due to the sublattice structure of graphene. Accounting for spin and orbital degrees of freedom implies that for nanotubes a shell structure is expected, where each shell can accommodate up to four electrons. In the absence of Coulomb interaction the energy levels are spin degenerate, while the orbital degeneracy is usually lifted due to the nanotube finite length. Coulomb interactions, however, modify this picture. The sublattice structure of graphene gives rise to a distinction between electron interactions on the same and on different sublattices. Therefore, besides the long-ranged forward scattering processes, also short-ranged interaction processes play a role in small diameter tubes \cite{Egg,Odin,Oreg,Leo1}. These short-ranged interactions cause in finite size nanotubes exchange effects leading for a tube filling of $4n+2$ to a groundstate with either total spin $S=0$ or $S=\hslash$ (a triplet)\cite{Leo1}. Signatures of the exchange interactions have indeed been inferred from stability diagrams of carbon-nanotube-based quantum dots \cite{Mor1,Sap,Liang}. In particular it was shown by Moriyama et al.\cite{Mor1} that an applied magnetic field can be used to reversibly change the groundstate from the singlet to one of the triplet states.
\newline Recently, carbon nanotubes have also attracted much attention for their potential applications in spintronic devices \cite{Cott}. They are particularly interesting because they  have a long spin lifetime and can be contacted with ferromagnetic materials. Indeed, spin-dependent transport in carbon nanotube spin valves has been demonstrated by various experimental groups \cite{Sahoo,Man,Haupt}, ranging from the Fabry-Perot \cite{Man,Sahoo} to the Kondo regime \cite{Haupt}.

From the theoretical point of view, spin-dependent transport in interacting SWNTs has been discussed so far in the limit of very long nanotubes \cite{Bal}, for tubes in the Fabry-Perot regime \cite{Peca} and for SWNT-based quantum dots \cite{Kos,Wey,Wey2}. In the three latter works the characteristic four-electron shell-filling could be observed in the stability diagrams. In \cite{Kos} however, focus was on medium-to-large diameter SWNTs where exchange effects can be neglected.
The studies in \cite{Wey,Wey2} are based on the theory by Oreg et al. \cite{Oreg}, where exchange interactions are treated on a mean-field level, and focus predominantly on shot noise\cite{Wey} and cotunneling\cite{Wey2} effects.
%In contrast to our work, their results are of purely numerical nature and no detailed analysis of the presented stability diagrams is performed. Actually, we will show that the low lying excitation lines are discriminative for the $S=0$ and $S=\hslash$ groundstate cases and can, for nonzero lead magnetization, reveal features like negative differential conductance.
\begin{figure}
\includegraphics[width=7cm]{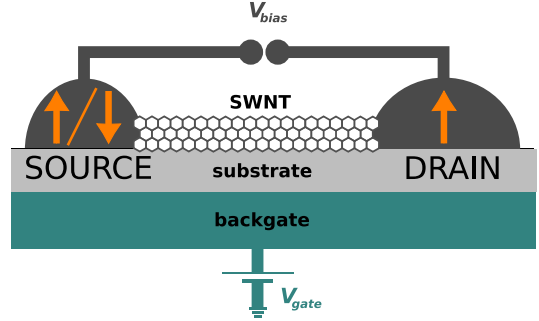}
\caption{\label{Fig:Setup} Single-electron-tunneling setup of a single-walled carbon nanotube (SWNT) which is weakly coupled to source and drain contacts. The contact magnetization may either be parallel or antiparallel as indicated by the arrows. The gate electrode allows to shift the chemical potential inside the SWNT.}
\end{figure}
\newline In this work we generalize the previous investigations of Ref.\cite{Kos} to include the short range Coulomb interactions causing exchange splittings of the six otherwise degenerate (at vanishing orbital mismatch) $4n+2$ - filling groundstates. The leads are either parallel or antiparallel spin-polarized and weakly coupled to the SWNT, see Fig. \ref{Fig:Setup}.
In the low bias regime we derive analytical formulas for the conductance for both large and small orbital mismatch corresponding to an $S=0$ and $S=\hslash$ groundstate, respectively, at $4n+2$ filling. In the high bias regime we numerically calculate the stability diagrams for  the two possible groundstates. We show several differences in transport between parallel and antiparallel lead magnetization, as e.g. a negative differential conductance (NDC) effect occurring only for the $S=0$ groundstate and antiparallel magnetization. We further include in the calculations a parallel magnetic field leading to a Zeeman splitting for all states with  total spin unequal to zero. It is then possible to observe spin blocking effects due to transport channels that trap the system in the triplet state with $S_z=-\hslash$. Performing a magnetic field sweep, a groundstate change may be obtained as it has been shown experimentally\cite{Mor1}.
\newline The paper is organized as follows. In section II we discuss the relevant features of the low energy Hamiltonian of interacting SWNTs with special focus on the filling $4n+2$.
%In this latter case the groundstate is either the triplet or a state with total spin zero, depending on the ratio between the exchange splitting $J$ and the orbital mismatch $\epsilon_{\Delta}$.
In section III we describe the set-up and method used to study spin-dependent transport in the sequential tunneling regime. Finally, in section IV, we present our results for the conductance, while in section V we focus on the nonlinear (finite bias) regime.

\section{The interacting low energy spectrum}\label{model}

%In the following we present the low energy Hamiltonian for the interacting carbon nanotube as it has been derived in chapter 2 of reference \cite{Leo1}. This Hamiltonian splits into a kinetic part, $H_0$, and two terms describing the interaction - a density part, $V_{\rho\rho}$, and a non-density part, $V_{n\rho\rho}$. While $H_0+V_{\rho\rho}$ can be diagonalized within the Tomonaga-Luttinger model we will calculate the contribution of $V_{n\rho\rho}$ in a truncated eigenbasis of $H_0+V_{\rho\rho}$. The reason for this procedure is that $V_{n\rho\rho}$ is no longer diagonalizable.

\subsection{The interacting Hamiltonian}\label{Hamiltonian}

The starting point for a microscopic, but still analytical, treatment of SWNTs is a tight-binding ansatz for the wavefunction of the $2p_z$ - electrons on the graphene honeycomb lattice. Including  nearest neighbor hopping matrix elements it yields an electron-hole symmetric bandstructure with a fully occupied valence band and an empty conduction band. Since the two bands touch at the cornerpoints of the 1st Brillouin zone, the Fermi-points, graphene is a zero gap semiconductor. Wrapping the considered sheet of graphene, i.e., imposing periodic boundary conditions (PBCs) around the circumference, yields a SWNT and leads to the formation of transverse subbands. For the low energy electronic structure of metallic SWNTs, only the subbands touching at the Fermi-points are of relevance. In the following we consider armchair SWNTs of finite length and impose open boundary conditions (OBCs) at the two ends of the tube, i.e., that the wave function vanishes at the armchair edges. This condition mixes the two inequivalent Fermi points $F=\pm K_0$ from the underlying graphene first Brillouin zone and yields the linear dispersion relation
\begin{figure}
\includegraphics[width=5cm]{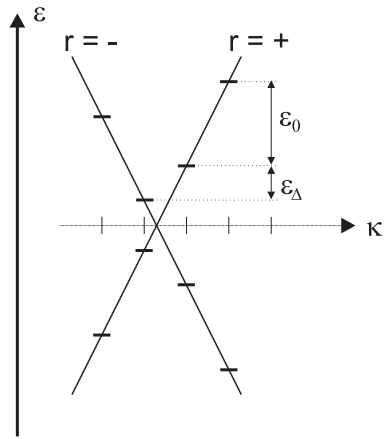}
\caption{\label{fig:BandstructureNanotube_OBC} The dispersion relation of a noninteracting SWNT with open boundary conditions. It is characterized by two linear branches, $r=\pm$, of slope $\pm \hbar v_F$ determined by the Fermi velocity $v_F$. The quantities $\epsilon_0$ and $\epsilon_{\Delta}$ are the intraband level spacing and the orbital mismatch energy, respectively. }
\end{figure}
of the finite size SWNT shown in Fig. \ref{fig:BandstructureNanotube_OBC}. It is characterized by two linear branches $r=\pm$ of slope $\pm \hbar v_F$ with the Fermi velocity $v_F\approx8.1\cdot 10^5\frac{m}{s}$.
 The allowed quasi-momentum values are given by $\kappa=(n_{\kappa}+\Delta)\pi/L$, where $n_{\kappa}\in\mathbb{Z}$, $L$ is the tube length and
$\Delta$ accounts for the fact that $K_0$ may not be an integer multiple of $\pi/L$.  The kinetic part of the Hamiltonian, yielding the energy relative to the Fermi-sea, correspondingly reads

\begin{equation}
\label{1}
H_{0}=\epsilon_0\sum_{r\sigma}r\sum_{n_{\kappa}}n_{\kappa}c^{\dagger}_{r\sigma\kappa}c_{r\sigma\kappa}+\epsilon_\Delta \sum_{r\sigma}rN_{r\sigma}\ ,
\end{equation}

\noindent where $\epsilon_0=\hslash v_F\pi/L$ is the level spacing,  and  $\epsilon_{\Delta}\equiv\epsilon_0\Delta$ is the  band offset energy.
Finally $c^{\dagger}_{r\sigma\kappa}$ creates an electron with momentum $\kappa$ and spin $\sigma$ in branch $r$ and the operator
$N_{r\sigma}$ counts the total electron number in branch $r$ and of spin $\sigma$.
\newline The interaction part of the Hamiltonian is given by

\begin{equation}
\label{2}
V\!=\!\frac{1}{2}\sum_{\sigma\sigma'}\!\int\!\!\!\int\! d^3\!rd^3\!r'\Psi^{\dagger}_{\sigma}(\vec{r})\Psi^{\dagger}_{\sigma'}(\vec{r}\,')U(\vec{r}-\vec{r}\,')\Psi_{\sigma'}(\vec{r}\,')\Psi_{\sigma}(\vec{r})\ ,
\end{equation}

\noindent where $\Psi,$ $\Psi^\dagger$ are fermion field operators and we use the Ohno-potential \cite{Barford},

\eq{146}{
U(\vec{r}-\vec{r}\,')=U_0 \left(1+\left(\frac{U_0\epsilon|\vec{r}-\vec{r}\,'|}{14.397}\right)^2\right)^{-\frac{1}{2}}eV\ \ ,
}

\noindent with $U_0=15\ \mbox{meV}$ \cite{Ful} and $\epsilon\simeq1.4-2.4$ \cite{Egg} is the dielectric constant of graphene. In the next step we express the 3D electron operators in terms of the 1D fermion-fields \cite{Leo2}

\begin{equation}
\label{3}
\psi_{rF\sigma}(x)=\frac{1}{\sqrt{2L}}\sum_{\kappa}e^{i\mbox{sgn}(F)\kappa x}c_{r\sigma\kappa}\ ,
\end{equation}
and obtain
\eq{101}{
\Psi_{\sigma}(\vec{r})=\sqrt{L}\sum_{rF}\mbox{sgn}(F)\psi_{rF\sigma}(x)\sum_p f_{pr}\varphi_{pF}(\vec{r})\ .
}

%\noindent it is possible to integrate over the coordinates perpendicular to the tube axis.
\noindent Here $F=\pm K_0$ denotes the two independent Fermi-points, $p=\pm$ the two sublattices of graphene, and the coefficients $f_{pr}$ of the sublattice wave function $\varphi_{pF}(\vec{r})$ are given by $1/\sqrt{2}$ for $p=+$ and $-r/\sqrt{2}$ for $p=-$. The sublattice wave function itself reads

\eq{102}{
\varphi_{pF}(\vec{r})=\frac{1}{\sqrt{N_L}}\sum_{\vec{R}\in L_G}e^{iFR_x}\chi_{p_z}(\vec{r}-\vec{R}-\vec{\tau}_p)\ ,
}

\noindent where $N_L$ is the number of graphene lattice sites identified by the lattice vector $\vec{R}$, and $L_G$ denotes the graphene honeycomb lattice in real space. Furthermore, $\chi_{p_z}(\vec{r}-\vec{R}-\vec{\tau}_p)$ is the $p_z$ wavefunction of a carbon atom living on sublattice $p$, identified by the sublattice vector $\vec{\tau}_p$. Upon integrating Eq. (\ref{2}) over the coordinates radial to the tube axis, one eventually arrives at a 1D interaction potential
characterized by density-density and non density-density contribution \cite{Leo1} so that the total Hamiltonian reads
\begin{equation}
\label{7}
H_{\odot}=H_0+V_{\rho\rho}+V_{{\rm n}\rho\rho}.
\end{equation}
With the help of bosonization \cite{Del} it is possible to diagonalize the density part  $H_0+V_{\rho\rho}$.
Eventually the bosonized and diagonalized Hamiltonian takes the form \cite{Leo1}:

\begin{multline}
\label{14}
H_0+V_{\rho\rho}=\sum_{j\delta q>0}\epsilon_{j\delta q}a^{\dagger}_{j\delta q}a_{j\delta q}+\frac{1}{2}E_cN^2_c\\
+\frac{1}{2}\sum_{r\sigma}N_{r\sigma}\left[-\frac{J}{2}N_{-r\sigma}+\left(\epsilon_0-u^+\right)N_{r\sigma}+r\epsilon_{\Delta}\right]\ \ .
\end{multline}
Besides the ground state, it accounts for all the possible fermionic and bosonic excitations of a SWNT.
\noindent The bosonic excitations are described by the first term on the right hand side. The indices refer to total/relative $(\delta=+/-)$ charge/spin $(j=c/s)$ modes. The energies $\epsilon_{j\delta q}$ are given by

\eq{186}{
\epsilon_{j\delta q}\cong\left\lbrace
\begin{array}{cc}
\epsilon_0n_q\sqrt{1+\frac{8W_q}{\epsilon_0}} & j\delta=c+\\
\epsilon_0 n_q& j\delta=c-,s+,s-
\end{array}
\right.\ ,
}

\noindent with $q=n_q\pi/L\ \mbox{for}\ n_q\in\mathbb{Z}$ and

\begin{multline}
\label{169}
W_{q}=\frac{1}{(2L)^2}\int_0^L dx\,\int_0^L dx' U^{\rm long}(x,x')\\
\times 4\cos(qx)\cos(qx')\ ,
\end{multline}

\noindent the contribution of the long-ranged density-density processes. Indeed $U^{\rm long}(x,x')=[U^{\rm intra}+U^{\rm inter}]/2$
is the sum of the interaction potentials for electrons living in the same (intra) and different (sublattices):
\begin{multline}
U^{\rm intra/inter}(x,x')=L^2\int\int d^2r_{\perp}d^2r'_{\perp}\\
\times\varphi^*_{pF}(\vec{r})\varphi^*_{\pm pF'}(\vec{r}\,')\varphi_{\pm pF'}(\vec{r}\,')\varphi_{pF}(\vec{r})U(\vec{r}-\vec{r}\,')\ .
\end{multline}
The second summand of (\ref{14})  is the charging term with the charging energy $E_c=W_{q=0}$ and also  comes from the long range part of the Coulomb interaction. It counts the energy one has to spend to put $N_c=\sum_{r\sigma}N_{r\sigma}$ electrons on the dot, no matter what spin $\sigma\in\{\uparrow,\downarrow\}$ or pseudospin $r\in\{+,-\}$ they have. The second line of (\ref{14}) starts with an exchange term favoring  spin alignment. The exchange-splitting,
\begin{multline}
\label{154}
J=\frac{1}{2N^2_L}\sum_{\vec{R} ,\vec{R}\,'}(1+e^{-i2K_0(R_x-R'_x)})\\
\times[U(\vec{R}-\vec{R}\,')-U(\vec{R}-\vec{R}\,'+\vec{\tau}_p-\vec{\tau}_{-p})]\ ,
\end{multline}
being proportional to the difference of the Coulomb interaction for electrons on the same and on different sublattices, accounts for the contribution of short range processes. The next term in (\ref{14}) reflects the energy cost for adding electrons of the same spin band in the same branch, i.e., the Pauli-principle, where the correction
$u^+$ is
\begin{multline}
\label{153}
u^+=\frac{1}{4N^2_L}\sum_{\vec{R} ,\vec{R}\,'}e^{-i2K_0(R_x-R'_x)}\\
\times[U(\vec{R}-\vec{R}\,')+U(\vec{R}-\vec{R}\,'+\vec{\tau}_p-\vec{\tau}_{-p})]\ .
\end{multline}
  Finally, the last term accounts for a possible band-mismatch, see Fig. \ref{fig:BandstructureNanotube_OBC}.
\newline The eigenstates of $H_0+V_{\rho\rho}$ are spanned by

\begin{equation}
\label{15}
\ket{\vec{N},\vec{m}}=\prod_{j\delta q}\frac{\left(\da_{j\delta q}\right)^{m_{j\delta q}}}{\sqrt{m_{j\delta q}!}}\ket{\vec{N},0}\ .
\end{equation}

\noindent Here $\vN\ \mbox{and}\ \vm$ denote the fermionic and the bosonic configuration, respectively, such that the state $\ket{\vec{N},0}$ has no bosonic excitation. The fermionic configuration is given by the number of electrons in each branch with a certain spin $\vN=(N_{-\uparrow},N_{-\downarrow},N_{+\uparrow},N_{+\downarrow})$. These eigenstates will be used to calculate the contribution of the non-density part of the interaction, i.e., $\bra{\vN,\vm}V_{{\rm n}\rho\rho}\ket{\vN',\vm'}$. Away from half-filling, they only couple states close in energy and one is allowed to work with a truncated eigenbasis (we check  convergence of the results as the basis is enlarged).
As shown by Yoshioka and Odintsov \cite{Yoshi}, for long SWNTs a Mott-insulating transition is expected to occur at half-filling due to umklapp scattering. As found in Ref. \cite{Leo1} umklapp processes acquire increasing weight as half-filling is approached also for finite size tubes, a possible signature of the Mott instability, and the present theory breaks down. In recent experiments \cite{Des} the observation of the Mott transition in SWNT quantum dots was claimed.
%\newline
%Introducing the bosonic operators
%\begin{eqnarray}
%\label{12}
%b_{r\sigma q}&:=&\frac{1}{\sqrt{n_q}}\sum_{\kappa}c^{\dagger}_{r\sigma\kappa}c_{r\sigma\kappa+q_r}\ ,\nonumber\\
%b^{\dagger}_{r\sigma q}&=&\frac{1}{\sqrt{n_q}}\sum_{\kappa}c^{\dagger}_{r\sigma\kappa}c_{r\sigma\kappa-q_r}\ ,
%\end{eqnarray}
%where $q_r:=r\cdot q$ for $q>0$ it is possible to rewrite the Hamiltonian in terms of collective particle-hole excitations. It is convenient to switch from the indices $r\sigma\longrightarrow j\delta$ referring to total (+) and relative (-) spin/charge modes instead of pseudospin and spin \cite{Leo1}. Moreover, it is necessary to perform a Bogoliubov transformation
%\begin{eqnarray}
%\label{13}
%a^{\dagger}_{j\delta q}&=&B_{j\delta q}b^{\dagger}_{j\delta q}+D_{j\delta q}b_{j\delta q}\ ,\nonumber\\
%a_{j\delta q}&=&B_{j\delta q}b_{j\delta q}+D_{j\delta q}b^{\dagger}_{j\delta q}\ ,
%\end{eqnarray}
%in order to finally obtain a Hamiltonian bilinear in the bosonic operators. The Bogoliubov-coefficients $B_{j\delta q}$ and $D_{j\delta q}$ have to be determined in a way that the new creators/annihilators fulfill the bosonic commutation relation.

\subsection{Low energy spectrum away from half-filling}

\begin{figure}
\includegraphics[width=7cm]{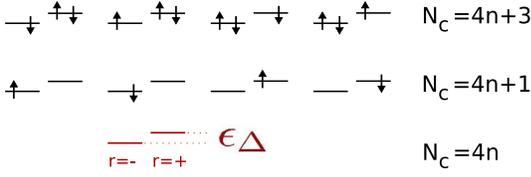}
\caption{\label{fig:Configuration} Lowest lying states for fillings $N_c=4n+1$ and $N_c=4n+3$. For simplicity only the configuration of the last partially filled shell is shown.}
\end{figure}

The low energy regime is where the energies that can be transferred to the system by the bias voltage and the temperature stay below $\epsilon_0$. This means no bosonic excitations are present, i.e., $\vm=(0,0,0,0)$, and also no fermionic excitations are allowed, i.e., the four bands will be filled as equal as possible: $|N_{r\sigma}-N_{r'\sigma'}|\leq1\, \forall\, r\sigma,r'\sigma'$. Our starting point are the eigenstates, Eq. (\ref{15}), of the Hamiltonian in Eq. (\ref{14}), which accounts for the kinetic and the density part of the full Hamiltonian. Now we have to split the examination into two cases.
\newline At first we consider states with total charge $N_c$ equal to $4n$, $4n+1$ and $4n+3$. Those are unambiguously described by the fermionic configuration $\vN$ because they are not mixed by the exchange effects.
The only impact of the short-range interaction terms on these states is given by
%the diagonal matrix element \cite{Leo1}
%
%\begin{equation}
%\label{16}
%\bra{\vN,0}V_{f^+bf^-}\ket{\vN,0}=u^+\sum_r\mbox{min}\left(N_{r\uparrow},N_{r\downarrow}\right)\ ,
%\end{equation}
%
%\noindent and leads to
an energy penalty for double occupation of one branch $r$, a common shift for all eigenstates with fixed $N_c\in \left\lbrace 4n,\ 4n+1,\ 4n+3\right\rbrace $. Therefore  we are left with \cite{Leo1}
\begin{multline}
\label{17}
E_{\vN}=\frac{1}{2}E_cN^2_c+u^+\sum_r\mbox{min}\left(N_{r\uparrow},N_{r\downarrow}\right)\\+\frac{1}{2}\sum_{r\sigma}N_{r\sigma}\left[-\frac{J}{2}N_{-r\sigma}+\left(\epsilon_0-u^+\right)N_{r\sigma}+r\epsilon_{\Delta}\right]
\end{multline}
\noindent for the energy. %This means that all possible groundstates for the different charge configurations in absence of a band-mismatch, i.e., $\epsilon_{\Delta}=0$, are energy degenerate.
If $\epsilon_{\Delta}\neq0$, states with the maximum allowed number of electrons in the $r=-$ branch will be the groundstates. For $N_c=4n$ the pseudospin branches $r=\pm$ are equally occupied, yielding an unique $N_c=4n$ groundstate. The corresponding configuration is taken as reference configuration for the $N_c=4n+1,\ 4n+2\ \mbox{and}\ 4n+3$ cases. The lowest lying states for $N_c\in \left\lbrace  4n+1,\ 4n+3\right\rbrace $ are presented in Fig. \ref{fig:Configuration}.
E.g., for the case $N_c=4n+1$ we obtain four possible states corresponding to $\vN\in\left\lbrace (n+1,n,n,n),\ (n,n+1,n,n),\ (n,n,n+1,n),\right.$ $\left.(n,n,n,n+1) \right\rbrace$. For simplicity we introduce for the states with an unpaired electron in the $r=-$ branch the notation $\ket{\uparrow,\cdot},\ \ket{\downarrow,\cdot}$. For electrons in the $r=+$ branch  we set $\ket{\cdot,\uparrow},\  \ket{\cdot,\downarrow}$.
%corresponding to one extra electron with spin $\uparrow/\downarrow$ and $r=-/+$, respectively.
%Since we will need the charge states $N_c=4n+1$ for the following transport calculations, we need to label them in a unique way. This will be done through the pseudospin-band index $r$ and spin-band index $\sigma$ and yields the four states $\ket{r\sigma}\in\left\lbrace \ket{-\uparrow},\ \ket{-\downarrow},\ \ket{+\uparrow},\  \ket{+\downarrow}\right\rbrace $.
\newline Analogously, neglecting exchange effects and setting $\epsilon_\Delta=0$ for the moment, the groundstates for the $N_c=4n+2$  filling are represented by the six states $\ket{\uparrow,\uparrow}$, $\ket{\downarrow,\downarrow}$, $\ket{\uparrow,\downarrow}$, $\ket{\downarrow,\uparrow}$, $\ket{\uparrow\downarrow,\cdot}$ and $\ket{\cdot,\uparrow\downarrow}$, where, e.g., $\ket{\uparrow,\uparrow}$ means two electrons with spin $\uparrow$ one on each branch $-$ and $+$.  Here the different fermionic configurations mix under the influence of the $V_{{\rm n}\rho\rho}$ processes and the groundstate structure will change dramatically due to off-diagonal contributions

\begin{eqnarray}
\label{18}
\bra{\uparrow,\downarrow}V_{{\rm n}\rho\rho}\ket{\downarrow,\uparrow}&=&-J/2\ ,\nonumber\\
\bra{\uparrow\downarrow,\cdot}V_{{\rm n}\rho\rho}\ket{\cdot,\uparrow\downarrow}&=&J/2\ .
\end{eqnarray}

%\noindent The energy of the lowest lying states is then given by
%
%\begin{small}
%\begin{multline}
%\label{19}
%E_{\vN}=E_{0,4n+2}\\\\
%+\left(
%\begin{array}{cccccc}
%-\frac{J}{2}&&&&&\\
%&-\frac{J}{2}&&&&\\
%&&0&-\frac{J}{2}&&\\
%&&-\frac{J}{2}&0&&\\
%&&&&u^+-\epsilon_{\Delta}&\frac{J}{2}\\
%&&&&\frac{J}{2}&u^++\epsilon_{\Delta}
%\end{array}
%\right)\ ,
%\end{multline}
%\end{small}

%\noindent where the matrix is represented in the above given basis.
\noindent Diagonalization of the interaction matrix yields the groundstate spectrum as it is shown in table \ref{States}. The energies in the table are given relative to $E_{0,4n+2}=\frac{1}{2}E_cN^2_c+(2n^2+2n+1)(\epsilon_0-u^+)-\frac{J}{2}(2n^2+2n)+2u^+n$.
\begin{table}
\begin{center}
\begin{tabular}{|c|}
\hline
 \begin{tabular}{|c|c|c|}
  \hline
   state & relative energy & spin \\
   \hline
   $\ket{t_1}=\ket{\uparrow,\uparrow}$ & $-J/2$ & $\hslash$\\
   $\ket{t_{-1}}=\ket{\downarrow,\downarrow}$ & $-J/2$ & $\hslash$\\
   $\ket{t_0}=\frac{1}{\sqrt{2}}\left(\ket{\uparrow,\downarrow}+\ket{\downarrow,\uparrow}\right)$ & $-J/2$ & $\hslash$\\
   $\ket{s}=\frac{1}{\sqrt{2}}\left(\ket{\uparrow,\downarrow}-\ket{\downarrow,\uparrow}\right)$ & $+J/2$ & 0\\
   $\ket{a}=\frac{1}{\sqrt{c^2_1+1}}\left(-c_1\ket{\uparrow\downarrow,\cdot}+\ket{\cdot,\uparrow\downarrow}\right)$ &       $u^+-\sqrt{\left(\frac{J}{2}\right)^2+\epsilon^2_{\Delta}}$ & 0\\
   $\ket{b}=\frac{1}{\sqrt{c^2_2+1}}\left(-c_2\ket{\uparrow\downarrow,\cdot}+\ket{\cdot,\uparrow\downarrow}\right)$ & $u^++\sqrt{\left(\frac{J}{2}\right)^2+\epsilon^2_{\Delta}}$ & 0\\
  \hline
 \end{tabular}\\
 \hline\\
 $c_1=\frac{2\epsilon_{\Delta}+\sqrt{J^2+(2\epsilon_{\Delta})^2}}{J},\quad c_2=\frac{2\epsilon_{\Delta}-\sqrt{J^2+(2\epsilon_{\Delta})^2}}{J}$ \\\\
 \hline
\end{tabular}
\end{center}
\caption{\label{States}The six lowest energy eigenstates for the filling $N_c=4n+2$ of an interacting SWNT. Due to short-ranged interactions there are three degenerate states of total spin $S=\hslash$ and three non-degenerate states of total spin $S=0$.}
\end{table}
It is clear that the states $\ket{s}$ and $\ket{b}$ will always be excited states, while the spin triplet, $S=\hslash$, is energy degenerate. Now the question arises which states, the triplet or the $\ket{a}$ state, are the groundstate of the system. In accordance with table \ref{States}, the condition for a triplet groundstate is given by:

\begin{eqnarray}
\label{21}
\epsilon^2_{\Delta}<(u^+)^2+Ju^+\ .
\end{eqnarray}

\noindent For a dielectric constant $\epsilon=1.4$ it holds
 $J=0.72\ \mbox{\AA{}}\ \frac{\epsilon_0}{d}$ and $u^+=0.22\ \mbox{\AA{}}\ \frac{\epsilon_0}{d}$. Hence we find in terms of the level spacing $\epsilon_0$ and the tube diameter $d$:

\eq{23}{
\left|\epsilon_{\Delta}\right|<0.4548\ \mbox{\AA{}}\ \frac{\epsilon_0}{d} .
}

\begin{figure}
\includegraphics[width=8.2cm]{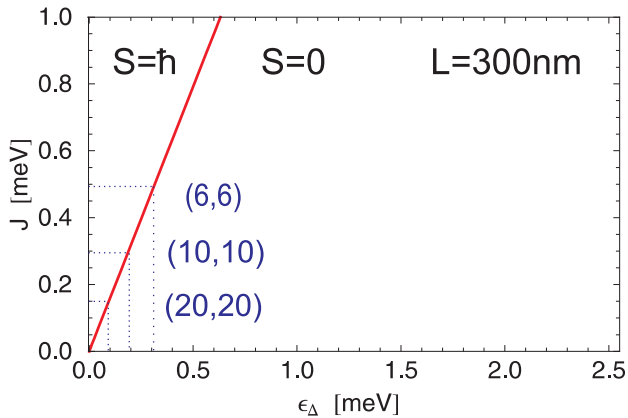}
\caption{\label{fig:Epsilon-J} Phase diagram to determine the groundstate of different tubes of length 300nm. The chance to find a triplet groundstate increases with increasing exchange parameter $J$, i.e., with decreasing tube diameters.}
\end{figure}

\noindent Obviously this makes the triplet groundstate more unlikely compared to the $S=0$ groundstate as it can be seen in Fig. \ref{fig:Epsilon-J}. For a (6,6) nanotube of 300nm length, the band-mismatch must be $\epsilon_{\Delta}<0.3\ \mbox{meV}\cong0.06\epsilon_0$ to be in a triplet groundstate. In the experiments \cite{Sap,Liang} band-mismatches are of the order of $0.3\epsilon_0$ and, as expected from our theory, $\ket{a}$ - groundstates are observed.

\section{Spin-dependent transport}\label{Transport}

%In this section we shortly outline how to derive the equation of motion (EOM) for the reduced density matrix (RDM) to lowest non-vanishing order in the tunneling (2nd order) Hamiltonian. We express the EOM for the RDM in the basis which diagonalizes the isolated system and, since we deal with spin polarized leads, in the spin quantization axis of the quantum dot. We will eventually show the formula for the current and the conductance in the low bias regime.

%\subsection{Equation of motion for the reduced density matrix}\label{EOM}

In this section we discuss  the set-up to evaluate spin-dependent transport across a SWNT weakly coupled to leads, see Fig. \ref{Fig:Setup}, and the main calculation tools.
The Hamiltonian of the full system reads

\eq{24}{
H=H_{\odot}+\sum_{l=s,d}H_l+H_T+H_{ext}\ ,
}

\noindent where $l=s,d$ denotes the Hamiltonian in the source and the drain contact, respectively. The leads magnetization is accounted for in terms of a Stoner Hamiltonian where the density of states, $\mathcal{D}_{l\sigma}(\epsilon)$, for the majority ($\sigma=\uparrow$) and the minority ($\sigma=\downarrow$) carriers are different. We treat the leads within the wide-band approximation,
 i.e., we regard the density of states as constant quantities  to be evaluated at the leads chemical potentials $\mu_s$ and $\mu_d$. We can thus define the polarization by ($l=s,d$):

\begin{eqnarray}
\label{51}
P_l=\frac{\mathcal{D}_{l\uparrow}(\mu_l)-\mathcal{D}_{l\downarrow}(\mu_l)}{\mathcal{D}_{l\uparrow}(\mu_l)+
\mathcal{D}_{l\downarrow}(\mu_l)} .
\end{eqnarray}
Moreover, we will consider a symmetric set up  $\mathcal{D}_{s\sigma}=\mathcal{D}_{d\sigma}=\mathcal{D}_{\sigma}$ and $P_s=P_d=P$. The total density of states is given by $\mathcal{D}_{tot}=\mathcal{D}_{\uparrow}+\mathcal{D}_{\downarrow}$.
We account for the bias voltage $V_b$ in terms of the difference $eV_b=\mu_s-\mu_d$ between the electrochemical potentials in the source and drain leads.
Further, $H_T$ in Eq. (\ref{24}) is the tunneling Hamiltonian which we will treat as a perturbation since  weak coupling to the leads is assumed.  Finally, $H_{ext}$ describes the influence of the externally applied gate voltage $V_g$.  The gate is capacitively coupled to the SWNT and hence contributes via a term $e\alpha V_gN_c$ with  $\alpha$  a proportionality factor.
\newline In order to evaluate the current-voltage characteristics we use the method developed in Ref. \cite{Kos} where, starting from the Liouville equation for the density matrix of the full system, a generalized master equation (GME) for the reduced density matrix $\rho$ (RDM) of the SWNT is obtained to second order in $H_T$.
 Once the stationary RDM is known, the stationary current through e.g. the source lead is evaluated from the relation
 $I_s=eTr\{\rho \dot N_s\}$, where $N_s$ is the number operator for electrons in the left lead.
 As this procedure with the relevant equations is thoroughly explained in Ref. \cite{Kos}, we refrain from repeating it here.
 % though the relevant equations are summarized in the Appendix.
 The GME can be solved in analytic form in the linear regime, being the focus of
 the following Sec. IV. In the nonlinear regime, discussed in Sec.  V,  the differential conductance is evaluated numerically.  Moreover, from here on we will focus on the transition between charge states $4n+1\longleftrightarrow4n+2$, mirror symmetric to $4n+2\longleftrightarrow4n+3$, as these two transitions are the ones that  reveal exchange effects. The remaining transitions $4n\longleftrightarrow4n+1$ and $4n+3\longleftrightarrow4(n+1)$ will not qualitatively change due to the presence of short range processes  and we hence refer  to the discussion  in \cite{Kos}. \\
 If not otherwise specified, we choose nanotubes described by the parameters in table \ref{Values}:
In order to obtain an $\ket{a}$ groundstate we assume a band-mismatch of $\epsilon_{\Delta}=0.3\epsilon_0=1.68\,$meV, whereas for a triplet groundstate we choose $\epsilon_{\Delta}=0$.

\section{The linear regime}

%In the low bias regime, we analytically solve Eq. (\ref{33}) in the stationary limit in order to obtain the populations of the groundstates. Once we know the populations we may subsequently calculate the current from Eq. (\ref{39}). Since we are in the low bias regime, we expand the formula for the current in linear order of the applied bias voltage such that the conductance is given by $G=\frac{\mbox{d}I}{\mbox{d}V}|_{V=0}=\frac{I}{V}$. For the actual calculation of the conductance in the low bias regime we choose a nanotube specified by the parameters given in table \ref{Values}.

\subsection{Conductance at zero magnetic field}

We focus on the conductance formulas for the two cases of tunneling from the $4n+1$ groundstates into the $S=0$ groundstate $\ket{a}$ or into the triplet groundstates.
\begin{table}
\begin{center}
 \begin{tabular}{|c|c|c|}
  \hline
    parameters & label & value \\
   \hline
   \hline
   length & $L$ & $300.06\,$nm\\
   diameter & $d$ & $0.81\,$nm\\
   dielectric constant & $\epsilon$ & $1.4\,$\\
   $\Downarrow$ & &  \\
   charging energy & $E_c$ & $6.7\,$meV\\
   level spacing & $\epsilon_0$ & $5.6\,$meV\\
   Coulomb excess energy & $u^+$ & $0.15\,$meV\\
   exchange energy & $J$ & $0.49\,$meV\\
   \hline
   orbital mismatch & $\epsilon_{\Delta}$ & $0\,$meV\ or\ $1.68\,$meV\\
  \hline
  thermal energy & $k_BT$ & $4.0\times10^{-3}\,$meV\\
  transmission coefficient & ${\cal D}_{\rm tot}\Phi$ & $1\times10^{-4}\,$meV\\
  \hline
 \end{tabular}
\end{center}
\caption{\label{Values} Parameter set of the 300nm (6,6) nanotube investigated in this work.}
\end{table}

For the transition $\ket{\sigma,\cdot}\longleftrightarrow\ket{a}$ the conductances in the case of parallel, $G^P$, and antiparallel, $G^{AP}$, magnetized leads are found to be
%\vspace{0.5cm}
%\underline{\textit{Singlet groundstate}}
%\vspace{0.5cm}

%\begin{figure}[]
%\begin{minipage}{5cm}
%\vspace{-0.9cm}
%\includegraphics[width=5cm]{graphics/singlet_0Grad_B0.eps}
%\caption{\label{fig:singlet_0Grad_B0} Conductance in the singlet groundstate for parallel magnetized leads. The lines denote analytical data whereas numerically calculated points are pictured by the dots.}
%\end{minipage}
%\hspace{0.35cm}
%\begin{minipage}{3cm}
%\vspace{0cm}
%\includegraphics[width=3cm]{graphics/singlet.eps}
%\caption{\label{fig:Transport_c1_parallel_B0} Transport for parallel lead magnetization. It is mediated by the majority electrons. Tuning up the polarization increases the majority current and correspondingly the minority current is diminished.}
%\end{minipage}
%\end{figure}

\begin{figure}
\includegraphics[width=\columnwidth]{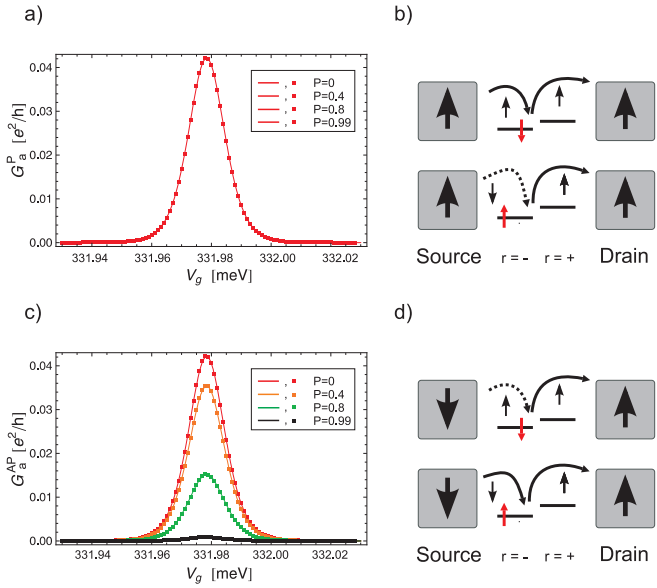}
\caption{\label{fig:singlet_0} Panels a) and c). Conductance vs. gate voltage for the $\ket{\sigma,\cdot}\longleftrightarrow\ket{a}$ resonance for parallel, $G^{P}_{a}$, and antiparallel, $G^{AP}_{a}$, lead magnetization. In both cases the analytical predictions Eqs. (\ref{40}), (\ref{41}) (continuous curves) perfectly match with the results from a numerical evaluation of the GME (squares). Strikingly $G^{P}_{a}$ is independent of the leads polarization $P$, while $G^{AP}_{a}$ is maximal at $P=0$.
Panels b) and d). Schematic explanation of the different polarization dependence.
The red spin specifies the spin of the state $\ket{\sigma,\cdot}$. The dashed/continuous arrows indicate rare/favorable tunneling processes. For parallel magnetization, panel b), the fast tunneling channel is the one with an excess spin $\downarrow$ and the electron transferred from source to drain is always a majority electron $\uparrow$. If the initial dot spin is $\uparrow$, this is likely to tunnel to the drain, such that at the end of the tunneling process a spin-flip has occurred, leaving the dot in the favorable configuration with a spin $\downarrow$.
For antiparallel lead magnetization, panel d), the fast channel corresponds to one electron in the dot with spin $\uparrow$. To this channel, however, is associated a spin flip. Because the situation with initial spin $\downarrow$ involves a rare tunneling process from the source lead, the conductance gets diminished
by increasing polarization.
}
\end{figure}

\begin{subequations}
\begin{align}
\label{40}
G^P_{a}&=\frac{c^2e^2\pi}{\hslash}\frac{\gamma}{1+\gamma}\ \beta\ \mathcal{D}_{tot}\ \Phi_{} \left \vert\frac{f(\mu_{a})f(-\mu_{a})}{2-f(\mu_{a})} \right \vert\ ,
\\\label{41}
G^{AP}_{a}
&=\frac{(P^2-1)\ \gamma\  (1+\gamma)}{P^2(\gamma-1)^2-(\gamma+1)^2}G^P_{a}\ ,
\end{align}
\end{subequations}

\noindent with $c={c_1}/{\sqrt{c^2_1+1}}$,
the Fermi function $f(\mu)$ evaluated at the gate voltage dependent energy difference $\mu_{a}=E_{\ket{a}}-E_{\ket{\sigma,\cdot}}$ and $\beta$ the inverse temperature.
 The parameters $\Phi=\Phi_s$ and $\gamma=\Phi_d/\Phi_s$ describe the possible asymmetric lead transparencies \cite{Kos} (hereby, $\Phi$ is in second order of the tunneling coupling contained in $H_T$).
The conductances
%plots belonging to Eqs. (\ref{40}) and (\ref{41})
are shown in Fig. \ref{fig:singlet_0}a) and \ref{fig:singlet_0}c) for the symmetric transparencies case $\gamma=1$ and ${\cal D}_{tot}\Phi=10^{-4}$meV. Strikingly, in the parallel magnetized case there is \textit{no} dependence on the polarization since there is never a blocking state involved in transport, see Fig. \ref{fig:singlet_0}b). For the antiparallel case, in contrast, transport is limited by the weakest channel (when there is a $\downarrow$ - electron on the dot) and one can drive the conductance to zero by tuning the polarization to $P\rightarrow1$. This feature is explained in Fig. \ref{fig:singlet_0}d).

%An interesting property is the current ratio, i.e., the tunnel magneto resistance (TMR), defined by

%\begin{eqnarray}
%\label{43}
% TMR=1-\frac{I_{N,N+1}(\theta=\pi)}{I_{N,N+1}(\theta=0)}\ ,
%\end{eqnarray}

%which in particular turns out to be independent of the gate voltage. We renounce to plot it, since it would only consist of distinct constant lines in the $V_G-\mbox{TMR}$-plane for different polarizations $P$ and asymmetries $\epsilon$ in the transparencies:

%\begin{eqnarray}
%\label{44}
% TMR=1-\frac{(P^2-1)(1+\epsilon)^2}{P^2(\epsilon-1)^2-(\epsilon+1)^2}\ .
%\end{eqnarray}
%\vspace{0.5cm}
%\underline{\textit{Triplet groundstate}}
%\vspace{0.5cm}
\begin{figure}
\includegraphics[width=\columnwidth]{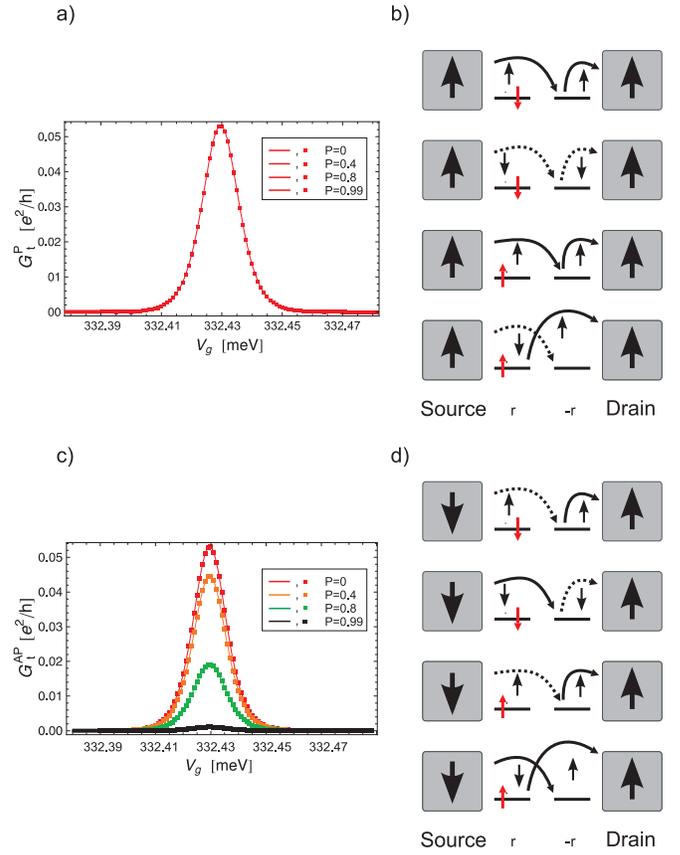}
\caption{\label{fig:triplet_0} Panels a) and c). Conductance vs. gate voltage at zero band-mismatch (triplet groundstate) for parallel, $G^{P}_{t}$, and antiparallel, $G^{AP}_{t}$, lead magnetization.  $G^{P}_{t}$ is independent of the leads polarization $P$, while $G^{AP}_{t}$ is maximal at $P=0$. The absolute value of the conductance is slightly larger than for the $ \ket{\sigma,\cdot}\longleftrightarrow\ket{a}$ case since more channels are involved. Panels b) and d). Schematic explanation of the different polarization dependence.
For simplicity we only drew the case in which the initial excess spin (red spin) is in the $r=+$ branch.
For parallel magnetization, panel b), the fast channel corresponds to the $\ket{\uparrow,\cdot}\longleftrightarrow\ket{t_{+1}}$ transition which conserves the spin of the excess dot electron. For antiparallel magnetization, panel d), the fast channel corresponds to an initial excess spin $\uparrow$ electron likely to tunnel to the drain and being replaced by a spin $\downarrow$ from the source. The situation with an initial spin $\downarrow$, however, corresponds to a weak channel.
 Increasing the polarization highly populates the $\ket{t_{-1}}$  state and transport decreases. }
\end{figure}

For the case of the triplet groundstate we face a completely new situation. First, we have for $N_c=4n+1$ filling four degenerate states available because the band-mismatch has been chosen to be zero. Secondly, we couple to three different states in the case of $N_c=4n+2$ rather than to just one. However, the conductance plots do not qualitatively change as it may be seen in Fig. \ref{fig:triplet_0}a) and \ref{fig:triplet_0}c).
%This is a somewhat sloppy procedure because for a band-mismatch $\epsilon_{\Delta}>k_BT$ the right (+) band would be unpopulated in the $4n+1$ - state like we had for the singlet groundstate.
The conductance formulas read:

\begin{subequations}
\begin{align}
\label{46}
G^P_{1,t}&=\frac{3e^2\pi}{\hslash}\frac{\gamma}{1+\gamma}\ \beta\ \mathcal{D}_{tot}\ \Phi\left \vert\frac{f(\mu_{t})f(-\mu_{t})}{4-f(\mu_{t})} \right \vert\ ,
\\
\label{47}
G^{AP}_{1,t}
&=\frac{(P^2-1)\ \gamma\  (1+\gamma)}{P^2(\gamma-1)^2-(\gamma+1)^2}G^{P}_{1,t}\ .
\end{align}
\end{subequations}

\noindent Compared to Eqs. (\ref{40}), (\ref{41}) the prefactor changed from $c^2$ to 3 due to the three involved triplet states. The quantity $\mu_{t}=E_{\ket{t}}-E_{1}$ is the difference between the triplet and the $N_c=4n+1$ - groundstate energies. In addition, the denominator in the term containing the Fermi-functions has also changed to account for the degeneracy of the $4n+1$ - filling states. The qualitative behavior, however, does not change compared to the case of an $\ket{a}$  groundstate, such that one cannot determine the spin nature of the groundstate from these plots alone.

\subsection{Conductance in the presence of an external magnetic field}

In this section we consider the influence of an externally applied magnetic field (Zeeman-field) which clearly reveals the character of the groundstate for $4n+2$ and, moreover, may even change the groundstate depending on the field strength. The field causes an additional Zeeman energy  to states with a spin-component $S_z\neq0$. The sign is negative if the concerned state in the tube is parallel to the external field and  positive if antiparallel. Thus, the chemical potential differences appearing in Eqs. (\ref{40}), (\ref{41}), (\ref{46}) and (\ref{47}) will be shifted by $\pm E_z=\pm \mu_{\rm B}B$. We use the convention $\mu_{\uparrow}=\mu-E_z$ and $\mu_{\downarrow}=\mu+E_z$. Furthermore, in order to improve the readability, we introduce the abbreviation $f_{\pm\uparrow/\downarrow}=f(\pm\mu_{\uparrow/\downarrow})$. % and we omit the indices $a$ and $t$ in the chemical potential.
The conductances for the antiparallel set-up are
\begin{subequations}
\begin{multline}
\label{52}
G^{AP}_{a}(E_z)=\frac{c^2e^2\pi}{2\hslash}\ \beta\ \mathcal{D}_{tot}\ \Phi\\
\times\left\vert\frac{f_{+\uparrow}f_{+\downarrow}(1+P(\gamma+1)+\gamma)f_{-\downarrow}}{f_{+\downarrow}+f_{+\uparrow}f_{-\downarrow}}\right.\\
\left. +\frac{f_{+\uparrow}f_{+\downarrow}(1-P(\gamma-1)+\gamma)f_{-\uparrow}}
{f_{+\downarrow}+f_{+\uparrow}f_{-\downarrow}} \right \vert\
\end{multline}
and
\begin{multline}
\label{53}
G^{AP}_{t}(E_z)=\frac{e^2\pi}{2\hslash}\ \beta\ \mathcal{D}_{tot}\ \Phi\\\times\Bigl|\Big\{ f_{-\uparrow}f_{-\downarrow}
\Bigl[\Bigl(1+\gamma-P(1-\gamma)\Bigr)f_{+\downarrow}\Bigl(f_{-\downarrow}f_{+\uparrow}+2f_{+\downarrow}f_{-\uparrow}\Bigr)\\
+\Bigl(1+\gamma+P(1-\gamma)\Bigr)f_{+\uparrow}\Bigl(f_{-\uparrow}f_{+\downarrow}+2f_{-\downarrow}f_{+\uparrow}\Bigr)\Bigr]\Big\}\Big/\\
\Bigl\{f_{-\uparrow}\Bigl( 1+f_{-\downarrow}\Bigr)\Bigl( f_{-\downarrow}f_{+\uparrow}+f_{-\uparrow}f_{+\downarrow}\Bigr)+f^2_{-\downarrow}f^2_{+\uparrow}\Bigr\}
\Bigr|\,.
\end{multline}\end{subequations}
\newline
We do not find qualitative differences with respect to the zero magnetic field case: the conductances decrease in both cases  with increasing polarization. In the following, we will therefore only focus on the parallel case, where we find interesting behavior for small Zemann splittings. The conductance formulas for parallel lead magnetization take the form
\begin{figure}
\includegraphics[width=\columnwidth]{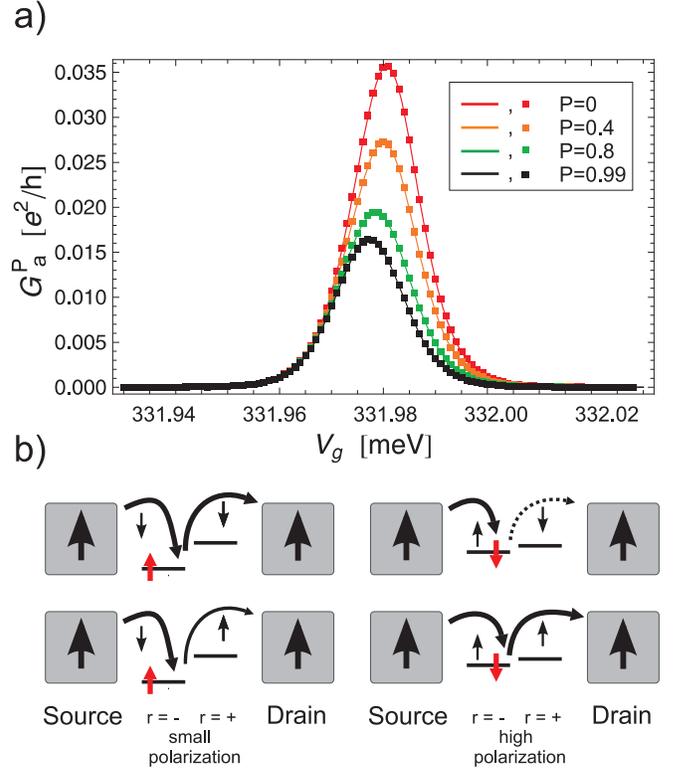}
\caption{\label{fig:singlet_007} a) Conductance near the $\ket{\sigma,\cdot}\longleftrightarrow\ket{a}$ transition for parallel magnetized leads and applied magnetic field. The peaks corresponding to  higher polarizations are shifted to lower gate voltages. b) Schematic explanation of the polarization  and gate-voltage dependence for small (left sketch) and large (right sketch) polarization.  The red spin indicates the spin of the excess electron initially present on the dot. The thick and thin lines are frequent and less frequent transitions, while dashed lines indicate rare transitions.
Large polarizations favor processes involving majority spins while, due to the extra required Zeeman energy, the Fermi function suppresses processes where a spin $\downarrow$ is transferred. Thus at small polarizations the transport is mostly mediated by spin $\downarrow$ - electrons while at large polarizations $\uparrow$ - electrons are preferred. Correspondingly the peak position is shifted to smaller gate voltages as the polarization is increased.
}
\end{figure}
\begin{subequations}
\begin{multline}
\label{48}
G^P_{a}(E_z)=\frac{c^2e^2\pi}{\hslash}\frac{\gamma}{1+\gamma}\ \beta\ \mathcal{D}_{tot}\ \Phi\\
\times\left\vert\frac{f_{+\uparrow}f_{+\downarrow}\Bigl[(P+1)f_{-\uparrow}-(P-1)f_{-\downarrow}\Bigr]}{f_{+\uparrow}f_{+\downarrow}+f_{+\downarrow}f_{-\uparrow}+f_{+\uparrow}f_{-\downarrow}} \right \vert\
\end{multline}
and
\begin{multline}
\label{49}
G^{P}_{t}(E_z)=\frac{e^2\pi}{2\hslash}\frac{\gamma}{1+\gamma}\ \beta\ \mathcal{D}_{tot}\ \Phi\\
\times\Bigl|
\Big\{ f_{-\uparrow}f_{-\downarrow}
\Bigl[(P+1)f_{+\uparrow}\Bigl(f^2_{-\uparrow}f_{+\downarrow}\\
+f_{+\uparrow}f_{-\uparrow}f_{+\downarrow}+2f_{+\uparrow}f_{-\uparrow}f_{-\downarrow}+2f^2_{+\uparrow}f_{-\downarrow}\Bigr)\\
-(P-1)f_{+\downarrow}\Bigl(f^2_{-\downarrow}f_{+\uparrow}+2f^2_{+\downarrow}f_{-\uparrow}\\ +2f_{+\downarrow}f_{-\uparrow}f_{-\downarrow}+f_{+\downarrow}f_{+\uparrow}f_{-\downarrow}\Bigr)\Bigr]\Big\}\Big/\\
\Bigl\{2f^2_{-\downarrow}f_{+\uparrow}f_{-\uparrow}+f^2_{+\uparrow}f^2_{-\uparrow}\\
+f^2_{-\uparrow}f^2_{+\downarrow}+2f^2_{-\uparrow}f_{-\downarrow}f_{+\uparrow}+f_{-\downarrow}f_{-\uparrow}f_{+\uparrow}f_{+\downarrow}\Bigr\}
\Bigr|\ .
\end{multline}\end{subequations}
\newline The corresponding plots can be seen in Figs. \ref{fig:singlet_007}a) and \ref{fig:triplet_007}a). In these calculations we considered a small magnetic field of $0.07\,$T which equals in magnitude the thermal energy of $k_BT=0.004\,$meV. This provides a situation with a finite occupation probability for all included states. Specifically, this means that also states containing $\downarrow$ - electrons will be populated, but the population of states containing $\uparrow$ - electrons will be preferred.
%\begin{figure}
%\includegraphics[width=7cm]{graphics/Energieschema_triplet.eps}
%\caption{\label{fig:Energiediagramm} Schematic drawing of how the energy levels for $4n+1$ and $4n+2$ (in the triplet groundstate) get split due to the magnetic field into three levels. Additionally, we show the transition lines and the associated chemical potential difference. It can be seen that the once degenerate transition line splits into two lines.}
%\end{figure}
 The first thing we observe in both Fig. \ref{fig:singlet_007}a) and \ref{fig:triplet_007}a) is that the once degenerate curves in Figs. \ref{fig:singlet_0}a) and \ref{fig:triplet_0}a) now split into distinct curves for the four different polarizations. Moreover, the peaks of the curves corresponding to less polarized leads continuously move to higher gate voltages. Finally the conductance \textit{decreases/increases} with increasing polarization for the $a/t$ cases, respectively. Let us examine the results starting with the $\ket{a}$ - groundstate.
%\vspace{0.5cm}
%\underline{\textit{Singlet groundstate}}
%\vspace{0.5cm}
We will divide the analysis in two cases, slightly polarized leads and strongly polarized leads.
\newline For only slightly polarized  or non-polarized leads the situation is intricate as we have to deal with \emph{competing processes}. On the one hand there is a highly populated $\ket{\uparrow,\cdot}$  state and a slightly populated $\ket{\downarrow,\cdot}$  state in the tube. From this point of view, the system prefers $\downarrow$ -  electrons to tunnel into the $\ket{a}$  state and to leave the dot subsequently such that the tube always remains in the preferred $\ket{\uparrow,\cdot}$  state (Fig. \ref{fig:singlet_007}, sketch b), upper left panel). Only rarely, the $\uparrow$ -  electron tunnels out, as this would result in a spin-flip to the disfavored $\ket{\downarrow,\cdot}$ state (Fig. \ref{fig:singlet_007}, sketch b), lower left panel).
\begin{figure}[]
\includegraphics[width=\columnwidth]{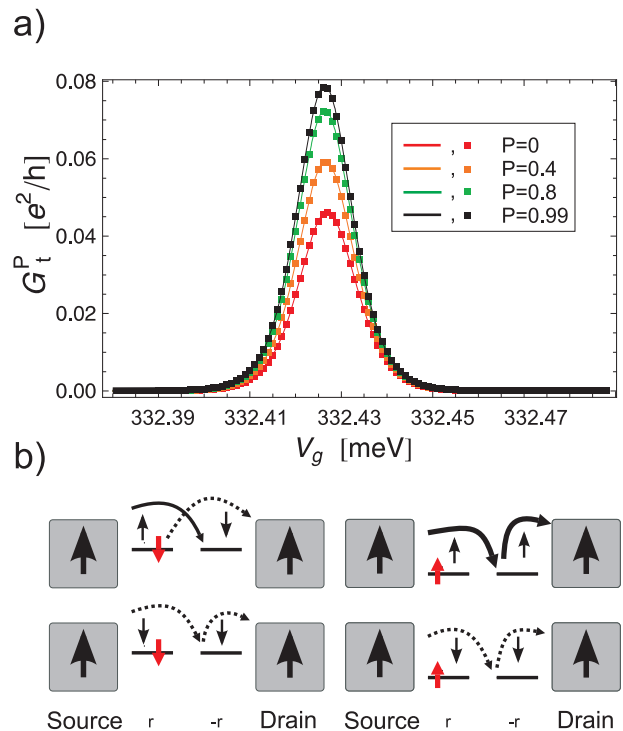}
\caption{\label{fig:triplet_007}  a) Conductance near the triplet resonance for parallel magnetized leads and applied magnetic field. In contrast to the case of a singlet resonance, Fig. \ref{fig:singlet_007}, transport increases as the  polarization is enhanced. b)Schematic explanation.  At small leads polarization the distribution of $\uparrow$ - electrons and $\downarrow$ - electrons is almost equal. However, the $\ket{t_1}$ - channel is preferred to the others. Increasing the polarization enhances the dominance of this channel  and correspondingly the conductance. Simultaneously the conductance peak is shifted to lower gate voltage indicating the dominance of $\uparrow$ - electrons.}
\end{figure} 
On the other hand, entering of $\downarrow$ - electrons is suppressed compared to transport of $\uparrow$ - electrons, not so much by the small polarization, but mainly due to the Zeeman splitting in the involved Fermi-functions: The chemical potential for  $\downarrow$ - electrons exceeds the one for  $\uparrow$ - electrons by $2E_z$ such that $f_{+\uparrow}>f_{+\downarrow}$ at any gate voltage. However, in the end it will be a mixture of mainly $\downarrow$ - electrons and some $\uparrow$ - electrons  responsible for transport. %The sketch b) of Fig. \ref{fig:singlet_007} describes this situation. 
This can also be seen by the fact that the curves for small polarizations are shifted to higher gate voltages which accounts for the higher chemical potential of the $\downarrow$ - electrons. In addition, the total amplitude of the conductance is decreased compared to the case without the magnetic field, Fig. \ref{fig:singlet_0}a), as there is always a limiting element - either the small Fermi-function or the small population - involved.
\newline In the case of highly polarized leads we face the situation where there are very few $\downarrow$ - electrons in the leads. As temperature provides a small, but nonzero population of the slightly excited state $\ket{\downarrow,\cdot}$, current mainly flows via the polarization-favored $\uparrow$ - electron channel. Since the chemical potential, the increment of the Fermi-functions, is smaller than in the former case the transition takes place at slightly lower gate voltages. The situation again is visualized in the sketch b) of Fig. \ref{fig:singlet_007}, in the upper and lower right panel.
%\vspace{0.5cm}
%\underline{\textit{Triplet groundstate}}
%\vspace{0.5cm}

At the triplet resonance we observe not only quantitative, but also qualitative changes. The plot can be seen in Fig. \ref{fig:triplet_007}a) and all relevant tunneling processes are sketched in Fig. \ref{fig:triplet_007}b). Let us again start with unpolarized or just slightly polarized leads. Due to a large population of the spin $\uparrow$ states in the $N_c=4n+1$ case and of the $\ket{t_1}$ state in the $N_c=4n+2$ case transport is mainly mediated via the majority charge carriers, i.e. $\uparrow$ - electrons (Fig. \ref{fig:triplet_007}b), upper right panel). %In addition, there is a big contribution from the Fermi-functions for transport of $\uparrow$ - electrons. 
However, the resulting current is smaller than in the case without magnetic field since it is harder to make use of the $\downarrow$ - electrons that are still largely at disposal in the leads. \newline A high polarization decreases the number of $\downarrow$ - electrons in the leads in favor of the $\uparrow$ - electron number, and such transport via the already preferred $\ket{t_1}$  channel is strongly enhanced.
%We may state:
%\begin{eqnarray}
%\label{50}
%G\propto\mathcal{D}_{\uparrow}f_{\uparrow}+\mathcal{D}_{\downarrow}f_{\downarrow}\qquad\mbox{with}\qquad f_{\uparrow}>f_{\downarrow}\ ,
%\end{eqnarray}
%and obviously the resulting conductance increases if we raise $\mathcal{D}_{\uparrow}$ while lowering $\mathcal{D}_{\downarrow}$.
As a consequence, the conductance by far \textit{exceeds} the conductance without magnetic field and polarization.
This effect should be detectable in an experimental setup and would give a possibility to distinguish between a triplet groundstate and a $S=0$ groundstate.

\section{The nonlinear regime}

\begin{figure}
\includegraphics[width=\columnwidth]{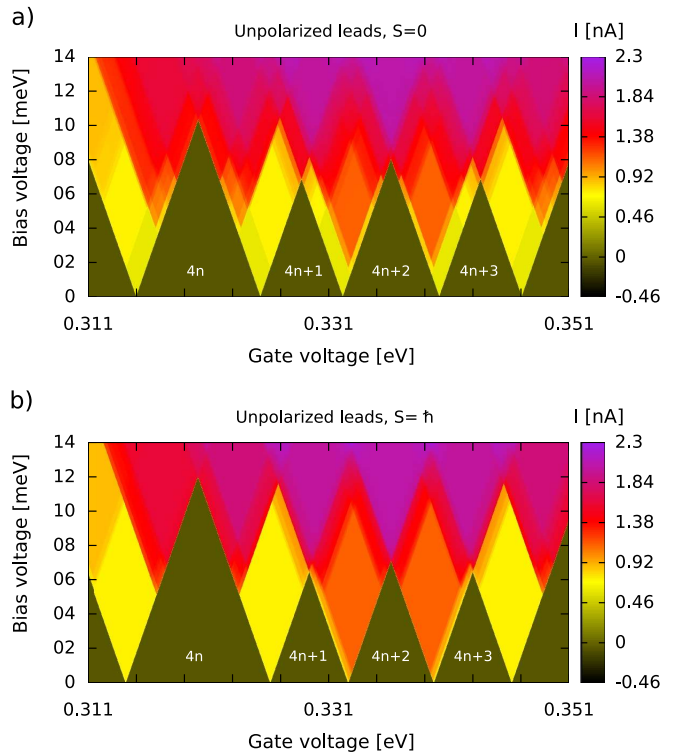}
\caption{\label{fig:CURRENT} Current versus gate and bias voltages for unpolarized leads. In total 176 states have been included, which corresponds to all states with at most one bosonic excitation. For $4n+2$-filling this amounts to 32 different states. a) Band-mismatch  $\epsilon_{\Delta}=0.3\ \epsilon_0$ corresponding to an $S=0$ groundstate for the $4n+2$ filling. b) Band-mismatch  $\epsilon_{\Delta}=0$ corresponding to an $S=\hslash$ groundstate at filling $4n+2$. In both cases a 4-electron periodicity of the Coulomb diamonds is observed.}
\end{figure}

In the finite bias regime also excited states become available and, due to the resulting high number of involved states, it is necessary to calculate the current numerically. We show the current and the stability diagrams - the differential conductance $\frac{\mbox{d}I}{\mbox{d}V_{b}}(V_b,V_g)$  as a function of the gate and the bias voltage.
%Additionally, we perform a magnetic field sweep - we show $G=\frac{\mbox{d}I}{\mbox{d}V_{g}}(B,V_g)$ - at a constant bias voltage.
The stability diagrams give a clear indication whether the involved groundstate in the transition $4n+1\longleftrightarrow 4n+2$ is the $\ket{a}$ state or the triplet. In the case of antiparallel lead magnetization we find negative differential conductance (NDC)  for transitions involving the $\ket{a}$  state. We also observe NDC  for transitions involving the $\ket{a}$ state or the triplet if an external magnetic field is applied.
%The magnetic field sweep reveals the in the experiment \cite{Mor1} measured groundstate change at a certain field strength.
\newline The current as a function of the gate and the bias voltage is shown in Fig. \ref{fig:CURRENT}a) for the $\ket{a}$ groundstate and in Fig. \ref{fig:CURRENT}b) for the triplet groundstate.
\vspace{0cm}
%Here we present the full pattern of diamonds which reveals a 2+2 electron periodicity caused by the exchange effects and the band-mismatch in the $\ket{a}$ - state and by the exchange effects only in the triplet.
All states with up to one bosonic excitation have been included in the calculation.
A 4-electron periodicity of the Coulomb diamonds is clearly seen.   The change in color indicates a change in current and therefore the opening of a new channel. At high bias a smearing of the transitions  due to the multitude of bosonic excitations is observed.
%Also, we detect Coulomb diamonds of different sizes, since for every fourth (4n) electron we have to pay the additional shell filling energy, and for the (4n+2) we have to deal with both exchange effects and the band-mismatch.
 In the remaining of this section we focus on the gate voltage region relevant for the $4n+1\longleftrightarrow4n+2$ transitions. In the plots of the differential conductance reported in the following we did not include the bosonic excitations  to avoid  a multitude of transition not relevant for the coming discussion. A polarization $P=0.9$ is chosen.

\subsection{Differential conductance at zero magnetic field}

\begin{figure}
\includegraphics[width=\columnwidth]{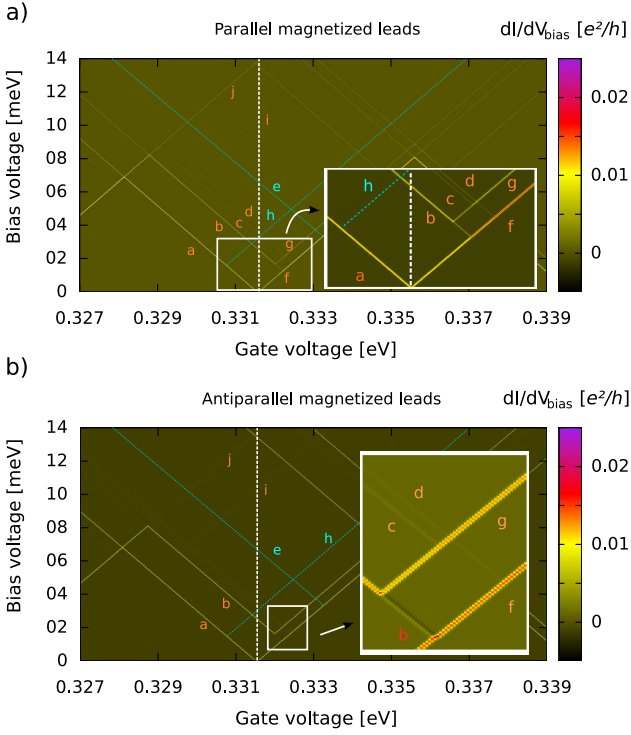}
\caption{\label{fig:singlet_Vg327-339_Vb0-14_B0} Differential conductance for transitions between $4n+1\longleftrightarrow4n+2$ filling in the $\ket{a}$ - groundstate. The polarization has been chosen to be $P=0.9$. The four lowest lying states for $4n+1$ and the six ones for $4n+2$ filling were included. The vertical white line is the bias trace we follow to explain the distinct transition lines in Fig. \ref{fig:Energiediagramm_c1_0_B0}. a) The leads are magnetized in parallel. b) Antiparallel magnetized leads. We observe a different intensity of the excitation lines between parallel and antiparallel magnetization. In particular a  pronounced negative differential conductance (NDC)  occurs in correspondence of the transition between $\ket{\sigma,\cdot}$ and the triplet (line $b$).}
\end{figure}

 Figs. \ref{fig:singlet_Vg327-339_Vb0-14_B0}a) and  Fig. \ref{fig:singlet_Vg327-339_Vb0-14_B0}b) show the stability diagrams for  parallel and antiparallel lead magnetization, respectively, for the case of the $\ket{a}$ groundstate. The two transition lines \textit{h} and \textit{e} were emphasized by a dashed line because these lines are so weak that it was not possible to resolve them together with the other stronger lines. The most obvious difference between the parallel and the antiparallel setup is the weakness of all transition lines beyond the triplet occupation (line \textit{b}) for antiparallel lead magnetization. Moreover an NDC line,  (line \textit{b}), not present in the parallel magnetization case, is observed.

\begin{figure}
\includegraphics[width=7cm]{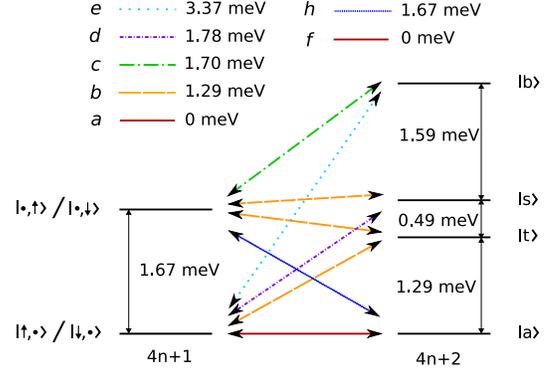}
\caption{\label{fig:Energiediagramm_c1_0_B0} Schematic drawing for the possible transitions occurring by sweeping the bias voltage  at the gate voltage that aligns the $\ket{\sigma,\cdot}$ and the $\ket{a}$-states (white dashed line in Fig. \ref{fig:singlet_Vg327-339_Vb0-14_B0}).}
\end{figure}

In order to explain the line positions in Fig. \ref{fig:singlet_Vg327-339_Vb0-14_B0}a),b) we provide a schematic drawing in Fig. \ref{fig:Energiediagramm_c1_0_B0} which is based on a bias trace at the particular gate voltage which aligns the groundstates (white vertical lines in Fig. \ref{fig:singlet_Vg327-339_Vb0-14_B0}). The differently colored arrows stand for new transport channels that open at certain bias voltages. The channels open in the order of \textit{a} to \textit{e} for transitions from $4n+1\longrightarrow4n+2$ (dashed arrows) and  \textit{f} to \textit{h} for transitions from $4n+2\longrightarrow4n+1$ (solid arrows). Sometimes opening of a new channel also opens other channels that have been blocked before and one does not see distinct lines for these transitions. Fig. \ref{fig:Energiediagramm_c1_0_B0} relates the concerned transitions to the required bias voltages.
Moreover, the line \textit{g}  stands for transitions between the triplet and the $\ket{\cdot,\sigma}$ states, i.e., it is a transition between excited states.
%Line e) indicates the transition from the (-) band to the $\ket{b}$ state at $eV_B/2=3.37\ \mbox{meV}$, whereas line h) stands for the transition from $\ket{a}$ to the (+) band at $1.67\ \mbox{meV}$.
%\newline Also we may have noticed the other two lines, i) and j), that stand for the transition from the (+) band to $4n$ and from the (-) band to $4n$, respectively.

\begin{figure}
\includegraphics[width=\columnwidth]{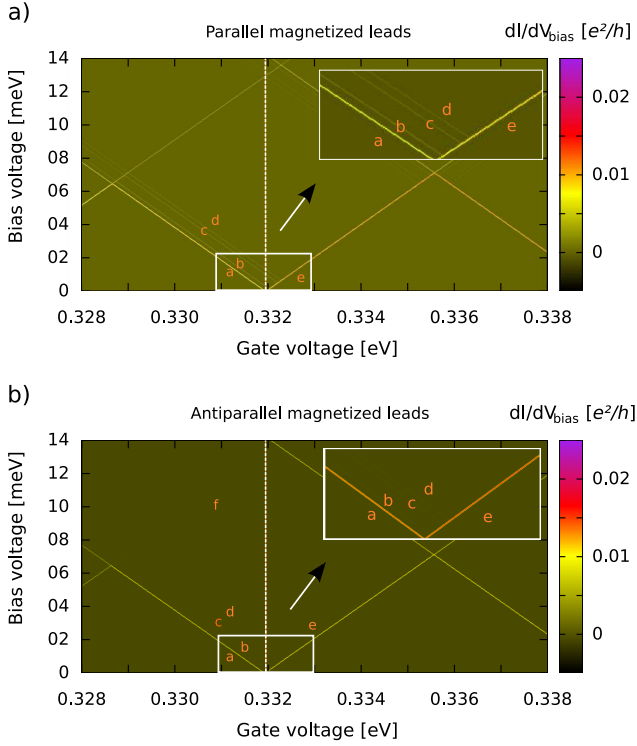}
\caption{\label{fig:trip_Vg328-338_Vb0-14_B0} Differential conductance for transitions between $4n+1\longleftrightarrow4n+2$ filling in the triplet  groundstate. The polarization has been chosen to be $P=0.9$. The four lowest lying states were included for $4n+1$ and the six lowest ones for $4n+2$. a) Leads parallel magnetized. b) Leads polarized antiparallel.  From the stability diagrams it is possible to directly extract the exchange parameters $u^+$ and $J$ since the bias voltage $V_b/2=u^+$ is needed to open transition line \textit{b} and $V_b/2=J$ to open line \textit{c}.}
\end{figure}

To explain the NDC in Fig. \ref{fig:singlet_Vg327-339_Vb0-14_B0}b) which follows upon line \emph{b} in the range between lines \textit{f} and line \textit{g}, we observe that --\,in correspondence of the \emph{b} line\,-- below the resonance only the transitions from $\ket{\sigma,\cdot}$ to the $\ket{a}$ state is possible. Above resonance also the triplet $\ket{t}$ is accessible. For the case of antiparallel polarization, both provide only weak transport channels: below the resonance
transport is mostly mediated by  $\uparrow$ - electrons (see also sketch of Fig. \ref{fig:singlet_0}) which are minority electrons for the source contact; above resonance, after some tunneling processes the system will always end up in the $\ket{t_{-1}}$ state which is a trapping state.
 Just at the exact resonance, the thermal energy allows electrons to tunnel forth and back, i.e., a $\downarrow$ - electron has the possibility to tunnel back into the source contact and transport is slightly enhanced. Once the bias voltage exceeds the exact resonance the trapping state $\ket{t_{-1}}$ gets occupied for long times and the current diminishes again. %The current does not completely drop to zero because  for these gate and bias voltages that open the transitions to the triplet, simultaneously, also the transition channels $\ket{t}\longleftrightarrow\left\lbrace \ket{\cdot,\uparrow},\ \ket{\cdot,\downarrow}\right\rbrace$ and $\left\lbrace \ket{\cdot,\uparrow},\ \ket{\cdot,\downarrow}\right\rbrace\longleftrightarrow\ket{s}$ open and escape from the trapping state is possible.
 %The escape from the trapping state becomes possible and NDC disappears.
\newline
%by the fact that it ends in the $\ket{\sigma,-}\longrightarrow\ket{t}$ line. Obviously, there cannot be a transition if the triplet state is not occupied.
%All other lines are symmetric to the just discussed ones, indicating transitions from $4n+2$ to $4n+3$. They do not reveal new physical aspects since we already know that transitions between $4n+1\longleftrightarrow4n+2$ can be treated in complete analogy to transitions between $4n+2\longleftrightarrow4n+3$.
%\newline For antiparallel lead magnetization, we obtain the stability diagram in Fig. \ref{fig:c1_180_B0}.
%The most obvious difference to the parallel case is the weakness of all transition lines beyond the triplet occupation (line b).
%\begin{figure}
%\includegraphics[width=7cm]{graphics/c1_180_B0.eps}
%\caption{\label{fig:c1_180_B0} Differential conductance for the $\ket{a}$ - groundstate in the antiparallel setup. The polarization has again been chosen to be $P=0.9$. For the explanation of the position of the lines Fig. \ref{fig:Energiediagramm_c1_B0} is still valid. However, the lines will change in their intensity due to the anti-parallel contact magnetization. The NDC-line (line b) for transitions from the (-) band to the triplet may clearly be seen in the inlet.}
%\end{figure}
The fact that the $\ket{\downarrow,\cdot}\longleftrightarrow\ket{t_{-1}}$ transition serves as the major transport channel  once it has been opened is also the
reason why all transition lines above line \textit{b} are so weak.
%Moreover, we observe in the inlet that the transition line from the (-) band to the triplet (line a) indicates a rise and immediate drop of current in the area
%
\newline In  Figs. \ref{fig:trip_Vg328-338_Vb0-14_B0}a) and \ref{fig:trip_Vg328-338_Vb0-14_B0}b) the stability diagrams for the $S=\hslash$ triplet groundstate are shown.
They look a lot simpler than the ones in Fig. \ref{fig:singlet_Vg327-339_Vb0-14_B0} due to the absence of a band-mismatch, causing a degeneracy of all four $4n+1$ filling  groundstates. Line \textit{a} is the groundstate to groundstate transition. Lines \textit{b} to \textit{d} indicate transitions from the $4n+1$  groundstates to $\ket{a}$, $\ket{s}$ and $\ket{b}$, respectively.
They come in the expected order, at an applied voltage $V_b/2$ equal to $u^+$, $J$ and $J+u^+$, as it is shown in table \ref{States}. Line \textit{e} stands for the transition from the triplet to one of the $4n+1$ groundstates.
\newline For the antiparallel setup, Fig. \ref{fig:trip_Vg328-338_Vb0-14_B0}b), we may see the same effect as we have observed in Fig. \ref{fig:singlet_Vg327-339_Vb0-14_B0}b), i.e., all lines beyond the transition to the triplet decrease in intensity.
%\begin{figure}
%\includegraphics[width=7cm]{graphics/trip_180_B0.eps}
%\caption{\label{fig:trip_180_B0} Differential conductance for the triplet groundstate in the anti-parallel setup. The polarization has been chosen to be $P=0.9$. All lines beyond the transition line to the triplet (line a) are too weak to be resolved.}
%\end{figure}
Since the triplet is the groundstate, this means all excitation lines are weak and may not be resolved in the figure.

\subsection{Differential conductance in parallel magnetic field\label{B}}

\begin{figure}
\includegraphics[width=\columnwidth]{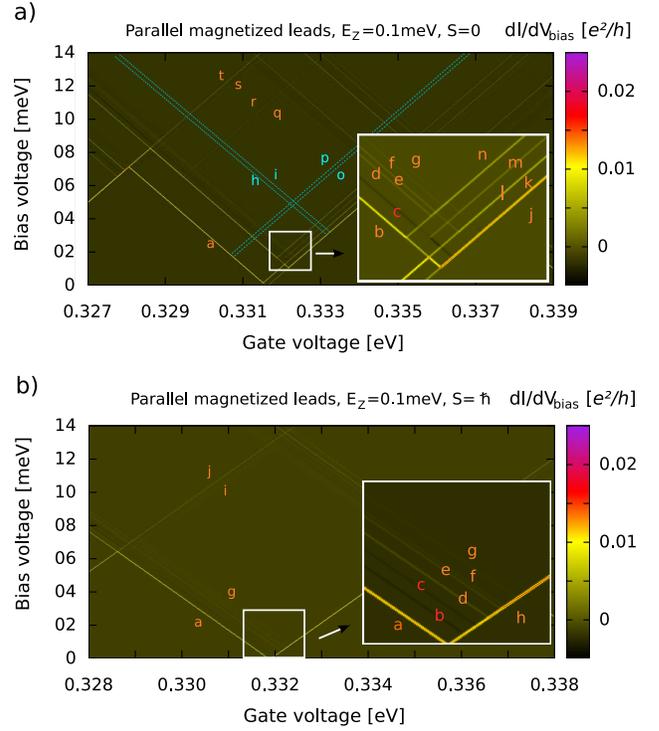}
\caption{\label{fig:c1_0_B01} Differential conductance for transitions between $4n+1\longleftrightarrow4n+2$ filling with an applied magnetic field of $E_z=0.1$ meV. A parallel lead magnetization was assumed with the polarization $P=0.6$. a) $\ket{a}$ -  groundstate.  Soon after line \textit{c} an NDC effect is observed due to the occupation of the $\ket{t_{-1}}$ trapping state. b) Triplet groundstate. After lines \textit{b} and \textit{c}  NDC occurs due to an increased population of the $\ket{t_{-1}}$ state.}
\end{figure}

Here we present results for an applied magnetic field of strength $E_z=0.1$ meV, Fig. \ref{fig:c1_0_B01}. The leads are parallel magnetized and a polarization of $P=0.6$ has been applied. The magnetic field removes  the spin degeneracy  of the triplet as well as of the $4n+1$ filled states; the resulting  Zeeman split transitions are clearly seen in Fig. \ref{fig:c1_0_B01} a) and are less well resolved in Fig. \ref{fig:c1_0_B01} b).

Explicitly, for the $\ket{a}$  groundstate,
%Line \textit{a} denotes the groundstate to groundstate transition $\ket{\uparrow,-}\longrightarrow\ket{a}$;
 line \textit{b} from Fig. \ref{fig:singlet_Vg327-339_Vb0-14_B0}  splits into lines \textit{b} and \textit{c} in Fig. \ref{fig:c1_0_B01}. We notice that line \textit{c} shows an NDC effect due to the opening of the channel $\ket{\downarrow,\cdot}\longrightarrow\ket{t_{-1}}$: though this transition, as mediated by minoriy $\downarrow$ - electrons, is rare, once it happens the system is trapped in the $\ket{t_{-1}}$ state for a long time due to the parallel polarization of the leads.
 % Moreover
%line \textit{c} from Fig. \ref{fig:singlet_Vg327-339_Vb0-14_B0} splits into lines \textit{d} and \textit{f} in Fig. \ref{fig:c1_0_B01};
%line \textit{d} from Fig. \ref{fig:singlet_Vg327-339_Vb0-14_B0} splits into lines \textit{e} and \textit{g} in Fig. \ref{fig:c1_0_B01};
 %line \textit{e} from Fig. \ref{fig:singlet_Vg327-339_Vb0-14_B0} splits into lines
%\textit{h} and \textit{i}, again emphasized by dashed lines.
For transitions from $4n+2$ to $4n+1$
%line \textit{f} from Fig. \ref{fig:singlet_Vg327-339_Vb0-14_B0} splits into lines \textit{j} and \textit{l}.
line \textit{k} is a new line that was Coulomb blocked in Fig. \ref{fig:singlet_Vg327-339_Vb0-14_B0}. It denotes the transition $\ket{s}\longrightarrow\ket{\cdot,\downarrow}$ and ends in line \textit{e} since the $\ket{s}$ state must be populated.
%Furthermore, there is line \textit{g} in Fig. \ref{fig:singlet_Vg327-339_Vb0-14_B0} that gets split into the lines \textit{m} and \textit{n}.
%The emphasized line \textit{h} from Fig. \ref{fig:singlet_Vg327-339_Vb0-14_B0} is now represented by the lines \textit{o} and \textit{p}.
Also, we  notice the absence of the $\ket{s}\longrightarrow\ket{\uparrow,\cdot}$ line since it is Coulomb blocked by the groundstate to groundstate transition  (line \textit{j}).
%As before the lines q) to t) represent the transitions from $4n+1$ to $4n$ which are now split by the magnetic field.

\begin{figure}
\includegraphics[width=7cm]{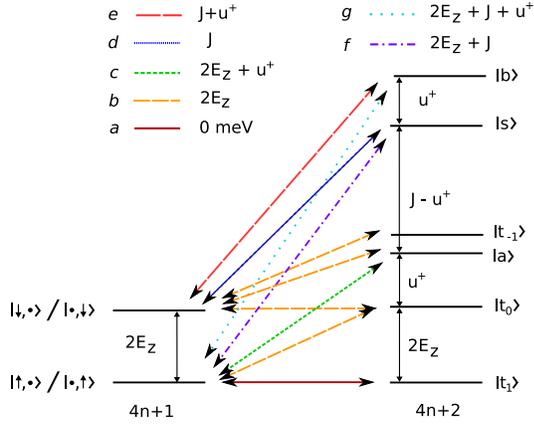}
\caption{\label{fig:Energiediagramm2} Schematic drawing of the possible transitions if the $\ket{\cdot,\uparrow}$ and $\ket{\uparrow,\cdot}$  states are aligned to the $\ket{t_1}$  state by the gate voltage at finite magnetic field and in the triplet groundstate. It provides the explanation for the transitions lines observed in the inset of Fig. \ref{fig:c1_0_B01}b).}
\end{figure}

 For the $S=\hslash$ triplet groundstate, Fig. \ref{fig:c1_0_B01}b), we observe that line \textit{b} and line \textit{c} show NDC effects. Line \textit{b} represents transitions from $\ket{\cdot,\uparrow}\longrightarrow\ket{t_0}$ or $\ket{\uparrow,\cdot}\longrightarrow\ket{t_0}$, which is  not  a trapping state. However, the applied bias voltage is sufficient to also populate the $\ket{\cdot,\downarrow}$ and $\ket{\downarrow,\cdot}$  states from $\ket{t_0}$ and subsequently from   $\ket{\cdot,\downarrow}$ and  $\ket{\downarrow,\cdot}$ the trapping state $ \ket{t_{-1}}$.  This process is also visualized in Fig. \ref{fig:Energiediagramm2}.
In the very same way it is possible to get trapped in the $\ket{t_{-1}}$ state via the $\ket{a}$ state indicated by line \textit{c}.
%Line a) is due to the groundstate to groundstate transition while line d) and e) denote transitions from the $\ket{\pm\downarrow}$ - state to the $\ket{s}$ singlet and the $\ket{b}$ - state, respectively. The lines f) and g) are caused by transitions from the $\ket{\pm\uparrow}$ - state to the $\ket{s}$ singlet and the $\ket{b}$ - state. Moreover, we observe that there are two lines - indicated by i) and j) - for the transition from $4n+1$ back to $4n$ because of the Zeeman split $4n+1$ - states.
%\begin{figure}
%\includegraphics[width=7cm]{graphics/trip_0_B01.eps}
%\caption{\label{fig:trip_0_B01} Differential conductance in the triplet groundstate with an applied Zeeman field of $E_z=0.1$ meV. The contact magnetization is parallel with $P=0.6$. Lines b) and c) show NDC effects due to an increased population of the $\ket{t_{-1}}$ trapping state.}
%\end{figure}

\subsection{The magnetic field sweep}

In a seminal experiment Moriyama et al.\cite{Mor1} demonstrated a transition from a  $S=0$  groundstate to a $S_z=\hbar$  groundstate upon magnetic field sweep in a SWNT quantum dot.
In this section we have computed the differential conductance in a gate-voltage and magnetic field plot both for unpolarized, as in  \cite{Mor1}, and parallel polarized leads with $P=0.9$.

We start from the $\ket{a}$  groundstate at $B=0$ with a band-mismatch of $0.24\ \epsilon_0$ (smaller than we previously used). This choice yields a change of groundstate from $\ket{a}$ to the triplet at a magnetic field  $\simeq6$ T as  measured experimentally\cite{Mor1}. To observe well visible patterns, we increased the temperature by a factor of ten compared to Tab. \ref{Values}.

%\begin{widetext}
\begin{figure}
\includegraphics[width=\columnwidth]{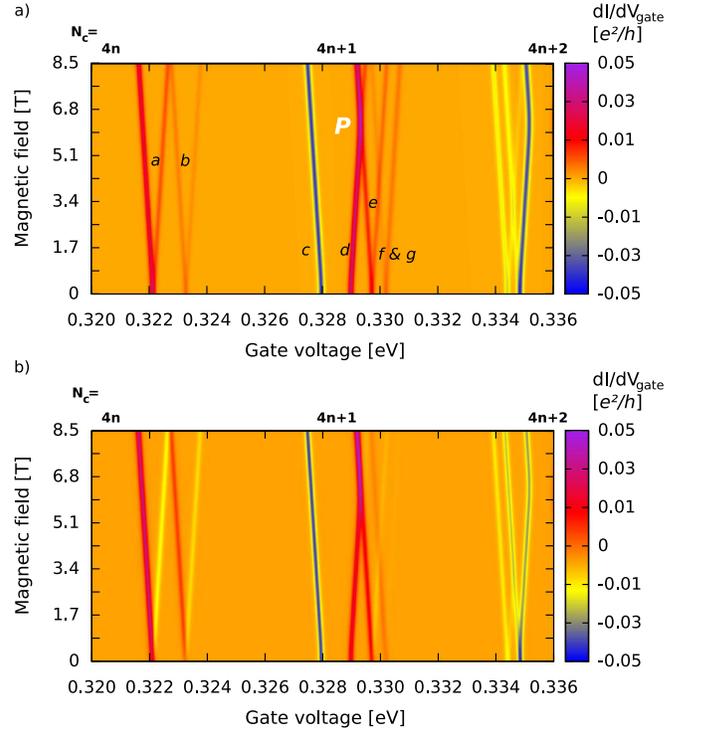}
\caption{\label{fig:singlet_Moriyama_Vg320-336_Vb0-14_B0-85} a) Differential conductance d$I$/d$V_g$ for a $B$-field sweep in the $\ket{a}$ - groundstate case. The applied bias voltage was fixed at $5.8\ \mbox{meV}$. Red lines indicate transitions that become possible at a certain gate voltage and blue lines show a transition that drops out of the transport window. The "V"-shaped patterns a and b represent transitions from $N_c=4n$ to $\ket{\sigma,\cdot}$  and $\ket{\cdot,\sigma}$, respectively. Each of the patterns is split by $2E_z$ denoting $\uparrow$ - electrons and $\downarrow$ - electrons tunneling in. At line $c$ we enter the $N_c=4n+1$ Coulomb diamond and transport gets suppressed. Line $d$ stands for the groundstate to groundstate transition from $\ket{\uparrow,\cdot}$ to the $\ket{a}$ - state. The "V"-shaped pattern $e$ is due to the transition $N_c=4n+1$ to the triplet whereas $f$ and $g$ denote transitions to the $\ket{s}$ - singlet and the $\ket{b}$  state, respectively. At the point $P$ the groundstate changes from the $\ket{a}$  state to the $\ket{t_1}$ - triplet. b) Ferromagnetic leads, polarized in parallel with $P=0.9$, are assumed. This changes the intensity of the transitions, while their positions are preserved. Moreover, transitions to excited states involving spin-down electrons are disfavored channels and hence converted from positive to negative differential conductance lines.}
\end{figure}
%\end{widetext}

The result of our calculation is presented in Fig. \ref{fig:singlet_Moriyama_Vg320-336_Vb0-14_B0-85}a). At a gate voltage of approximately $0.322\ \mbox{meV}$ and $0.323\ \mbox{meV}$ we have two $V$-shaped transition patterns (\textit{a} and \textit{b}) each of width $2E_z=2\mu_BB$. The separation between $a$ and $b$ at zero field is the band-mismatch $\epsilon_\Delta$.
%This is due to the fact that the energy of the $(\ket{\uparrow,-},\ \ket{-,\uparrow})$ - states is lowered by $\mu_BB$, while the energy of the $(\ket{\downarrow,-},\ \ket{-,\downarrow})$ - states is increased by the same amount.
Interestingly, for polarized leads, the branches belonging to transitions involving $(\ket{\downarrow,\cdot},\ \ket{\cdot,\downarrow})$, corresponding to the positive slope of the $"V"$, are NDC lines, Fig. \ref{fig:singlet_Moriyama_Vg320-336_Vb0-14_B0-85}b). The reason is the same as addressed already in section \ref{B}: once the $\downarrow$ - channel becomes available, there is some chance that from time to time a minority charge carrier ($\downarrow$ - electron) enters from the source. As the drain is polarized in parallel to the source, it will take quite a while until this electron can leave the SWNT again, such that transport gets hindered. At the gate voltage of  approximately $0.328\ \mbox{meV}$, one enters the $N_c=4n+1$ Coulomb diamond (line \textit{c}) and transport gets completely suppressed. The dot is in the groundstate  $ \ket{\uparrow,\cdot}$ at $B\neq0$. At $V_g\simeq0.329\ \mbox{meV}$ transport from $N_c=4n+1$ to the $\ket{a}$  state is enabled (line \textit{d}).
%The reason that we only discover one line, which is of positive differential conductance both for polarized and for unpolarized leads, is the following: The first possible transition is the groundstate to groundstate transition, i.e., from $\ket{\uparrow,-}$ to $\ket{a}$. Increasing the gate voltage further also groundstate to excited state and excited state to excited state transitions are possible. However, a transition from an excited state to the groundstate cannot be observed since it would occur at a lower gate voltage than the groundstate to groundstate transition, i.e., it is Coulomb blocked. Hence we only see one line --\, with positive slope\, -- indicating the transition from $\ket{\uparrow,-}$ to $\ket{a}$.

The next transitions (patterns \textit{e}, \textit{f}, \textit{g}) we observe are again split by $2E_z$ and therefore shaped like a "$V$". In all cases, the positively sloped branches are now again of NDC nature for a parallel lead polarization. The first "$V$" belongs to the triplet (pattern \textit{e}) and is of stronger intensity than the following two patterns. The transitions $\ket{\uparrow,\cdot}\longleftrightarrow\ket{t_1}$ and $\ket{\downarrow,\cdot}\longleftrightarrow\ket{t_0}$
contribute to the negative sloped part, while $\ket{\uparrow,\cdot}\longleftrightarrow\ket{t_0}$ and $\ket{\downarrow,\cdot}\longleftrightarrow\ket{t_{-1}}$  are responsible for the positive shaped line.
 The crossing of the $e$ and $d$ lines occurring at $B\cong6\ \mbox{T}$, point $P$, indicates \emph{the change in the groundstate from $\ket{a}$ to the state  $\ket{t_1}$.}

From the triplet pattern \textit{e} the additional gate voltage equal to the exchange energy  $J$ is needed to arrive at the last two "$V$" - shaped patterns $f$ and $g$. Compared to the lines for the triplet transition they are quite close to each other and of less intensity. These lines belong to a transition from both the $\ket{\downarrow,\cdot}$ and the $\ket{\uparrow,\cdot}$  states to the $\ket{s}$ - singlet (pattern \textit{f})  and the $\ket{b}$  state (pattern \textit{g}).  Finally, the lines on the right edges of the plots are mirror images and belong to backward transitions from $N_c=4n+2$ to $N_c=4n+1$; for this reason they mark a decrease of current for both polarized and unpolarized leads.

\section{Conclusions}

In summary, we have calculated spin dependent transport through fully interacting SWNTs in both the linear  and the nonlinear regime, with and without an applied magnetic field.

Peculiar of metallic SWNTs of small diameter is the possibility, due to exchange interactions, to find the system at $4n+2$ filling either in a groundstate of total spin $S=0$  or $S=\hbar$. Which of the two groundstates occurs in a real nanotube depends on the relation between the exchange energy and the orbital band mismatch. Thus, with focus on transitions involving $4n+1 \longleftrightarrow 4n+2 $ filling,  we investigated both situtations and demonstrated pronounced differences in the current-voltage characteristics depending on the considered groundstate.
\newline For example in the  linear regime the conductance for  parallel lead magnetization and finite magnetic field increases by raising the polarization for the case of a triplet groundstate but it decreases for the $S=0$ groundstate. This is due to the fact that for the triplet groundstate transport is dominated by a channel involving the triplet state $\ket{t_1}$ (with both spins $\uparrow$);  for the $S=0$ case  transport to be mediated by the majority electrons requires to make use of the $4n+1$ lowest excited state $\ket{\downarrow,\cdot}$ (and hence less favorable), Zeeman split from the ground state.

In the nonlinear regime we presented stability diagrams with parallel and antiparallel lead magnetization for both ground sates. In the antiparallel case it was possible to observe a negative differential conductance (NDC) effect for the $S=0$ groundstate, following immediately upon a conductance enhancement at the opening of a trapping channel to the excited triplet state $\ket{t_{-1}}$. Directly at that resonance, electrons can, just by thermal activation, tunnel back \emph{and} fourth, such that trapping in the $\ket{t_{-1}}$ state can not yet act, leading to an intermediate conductance increase. Away from resonance, the blocking effect fully occurs, resulting in the NDC. By adding an external magnetic field in the parallel setup we found NDC effects for both groundstates caused by spin blocking mediated by $\downarrow$ - channels, involving in particular the triplet state $\ket{t_{-1}}$.

 Finally,  we also presented results for the differential conductance in a gate-voltage and magnetic field map at finite bias.  These  magnetic field sweeps immediately allow to recognize the nature of the $4n+2$-filling groundstate at zero field, as well as to tune the nature of the groundstate from $S=0$ to $  S_z=\hbar$ upon variation in the field amplitude.
 Our results for unpolarized leads are in \emph{quantitative} agreement
  with experiments on a small-diameter SWNT by Moryama et al. \cite{Mor1}.
  Importantly the sweep at zero field also allows to immediately read off the values of the short range interactions $J$ and $u^+$. Specifically,   $J$ is the singlet-triplet exchange splitting and $u^+$ characterizes at zero orbital  mismatch  the energy difference between two  of the low energy states of total spin $S=0$. In the presence of polarized leads the magnetic field sweep also reveals lines of NDC due to the trapping nature of all $\downarrow$ - channels.

  The predictions of our theory are in quantitative agreement with experimental results obtained so far for
  unpolarized leads \cite{Mor1,Sap,Liang}. Due to recent achievements on spin-polarized transport in SWNTs \cite{Sahoo,Man,Haupt},  our predictions on spin-dependent transport are within the reach of present experiments.

\section{acknowledgments}

We acknowledge support by the DFG under the funding programs SFB 689, GRK 638.

\end{document}